**A Personalized Fluid-structure Interaction Modeling Paradigm for Aorta in Human Fetuses**

Zhenglun Alan Wei[1,*], PhD; Guihong Chen[2,*], MD; Biao Si[1], MD, PhD; Liqun Sun[3], MD PhD; Mike Seed[3], MD; and Shuping Ge[4], MD

[1]Department of Biomedical Engineering, Worcester Polytechnic Institute, Worcester, MA, USA

[2]Medical Ultrasound Department, The Fourth Hospital of Shijiazhuang, Shijiazhuang, China

[3]Division of Cardiology, Department of Pediatrics, The Hospital for Sick Children, University of Toronto, Toronto, Canada

[4]Department of Pediatric and Adult Congenital Cardiology, Geisinger Heart and Vascular Institute, Geisinger Clinic, Danville, PA, USA

* This work is a result of the equal contributions of the following authors.

**Please send the correspondence of the current work to**

Name: Zhenglun Alan Wei, PhD
Address: 60 Prescott Street
Phone: 978-934-3754
Fax:
E-mail: ZWei1@wpi.edu

Name: Shuping Ge, MD, FACC, FAHA, FASE
Address: 100 N. Academy Avenue, Danville, PA  17822
Phone: 570-271-6466
Fax: 570-214-1480
E-mail: sge@geisinger.edu

**Abstract (Word Count: 232; Limit: 250)**

Fluid-structure interaction (FSI) modeling, a technique widely used to enhance imaging modalities for adult and pediatric heart diseases, has been *underutilized in the context of fetal circulation because of limited data* on flow conditions and material properties. Recognizing the significant impact of congenital heart diseases on the fetal aorta, our research aims to address this gap by developing and validating a personalized FSI model for the fetal aorta.

Our approach involved reconstructing the anatomy and flow of the fetal aorta using fetal echocardiography and ultrasound. We developed an innovative iterative method that includes: (i) an automated process for incorporating Windkessel models at outflow boundaries when clinical data is limited because of the resolution constraints of fetal imaging, (ii) an inverse approach to estimate bulk material properties, and (iii) an FSI model for high-fidelity hemodynamic evaluation. This method is efficient, typically converging in fewer than three iterations.

We analyzed four normal fetal aortas with gestational ages ranging from 23.5 to 35.5 weeks to validate our workflow. We compared results with *in vivo* velocity waveforms across a cardiac cycle at the aortic isthmus. Strong correlations ($R>0.95$) were observed. Furthermore, our findings suggest that the stiffness of the fetal aorta increases until 30 weeks of gestation and then decreases.

This study marks a *first-of-its-kind* effort in developing a rigorously validated, personalized flow model for fetal circulation, offering novel insights into fetal aortic development and growth.

ABBREVIATIONS

| | |
|---|---|
| 2D | two-dimensional |
| 3D | three-dimensional |
| AAo | ascending aorta |
| AoA | aortic arch |
| PA | pulmonary artery |
| ThAo | thoracic aorta. |
| AoI | aortic isthmus |
| BA | brachiocephalic artery |
| CHD | congenital heart disease |
| DA | ductus arteriosus |
| DAo | descending aorta |
| FE | fetal echocardiography |
| FEA | finite element analysis |
| FSI | fluid-structure interaction |
| GA | gestational age |
| LCCA | left common carotid artery |
| LSA | left subclavian artery |
| PyAuto3WK | in-house Python code |
| R | distal resistance |
| WK | Windkessel |
| WSS | wall shear stress |
| Z | proximal resistance |

## A. Introduction

Congenital heart disease (CHD) is the most common congenital disability and a leading cause of death and chronic disease in newborns, infants, and children [1-3]. Prenatal screening and identification of CHD provide the earliest window of opportunity to improve the understanding of its pathogenesis, timely diagnosis/prognosis, parental counseling, safety, and effective therapy for optimal perinatal and long-term outcomes [4, 5]. However, there are numerous challenges in clinical care and innovation in CHD due to the lack of animal models, its rarity, and limited resolutions of contemporary fetal imaging methods for the accurate prenatal diagnosis and prognosis of CHD [6, 7].

Computational flow modeling has been used to augment medical imaging modalities and capture high-fidelity hemodynamics in the cardiovascular system in a personalized fashion. Many previous studies have used computational models to understand and uncover the mechanism, improve diagnosis/prognosis, and develop treatment methods such as medical devices and therapies for adult and pediatric heart diseases [8-11]. However, data for developing a computational model for fetal circulation are scarce. *Few models have been rigorously validated* against clinical measurements. Wiputra *et al.* developed a computational model to simulate the fetal right ventricle and obtained "generally good agreements" between their simulated data and a range of published data [12]. Later, the same group presented a plausible study to validate their model against clinically measured velocity by echocardiography. However, their validation focused on the velocity at specific time points of a cardiac cycle such as, at peak systole, instead of time-resolved quantities over the whole cardiac cycle [13]. Last year, Salman *et al.* used the model from Wiputra's study and confirmed that the model produces results that qualitatively agree with the data found in the literature [14].

The fetal aorta is involved in many CHDs, such as coarctation of the aorta (CoA), patent ductus arteriosus (DA), and transposition of the great arteries. Researchers have developed many computational models for the aorta in pediatric or adult patients [15-20]. However, the fetal aorta is NOT a smaller version of the pediatric or the adult aorta; both of which have an additional vessel (the DA) (Figure 1) that makes the anatomy and flow conditions more complex.

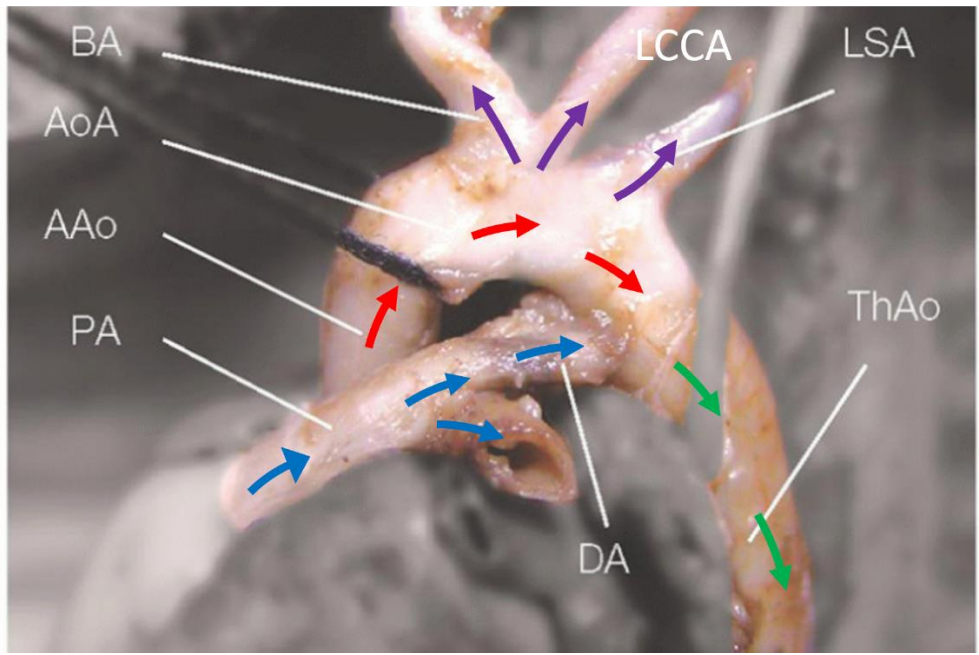

Figure 1. An illustration of the fetal aorta in a healthy individual [21]. AAo: ascending aorta; AoA: aortic arch; BA: brachiocephalic artery; DA: ductus arteriosus; LCCA: left common carotid artery; LSA: left subclavian artery; PA: pulmonary artery; and ThAo: thoracic aorta.

Previous studies used zero-dimensional lumped parameter models that simulate the hemodynamics in the fetal aorta using a simplified relation between flow rate and pressure drop [22-25]. Recent studies used computational fluid dynamics models to capture high-fidelity three-dimensional (3D) hemodynamics for the fetal aorta [26-28]. However, these models overlook many essential patient characteristics such as, the material properties of the vessel, which could lead to significant errors in calculating hemodynamic metrics [15, 19, 20, 29]. More importantly, these models have yet to be validated. Therefore, to date, there is *no rigorously validated computational model* for the fetal aorta.

Thus, this study aimed to develop a novel, personalized flow model to assess the hemodynamics of the fetal aorta and validate it using *in vivo* measurements. The developed model involves personalized anatomy, flow, and material properties of the fetal aorta. The validated model can be used to understand the pathogenesis of CHD related to the fetal aorta, improve pertinent diagnostic/prognostic methods, and investigate the feasibility, safety, and efficacy of therapeutic strategies.

## B. Materials and Methods

The process of developing the fluid-structure interaction (FSI) model begins with the anatomical reconstruction of the fetal aorta, utilizing images and data obtained from dynamic 3D SpatioTemporal Image Correlation fetal echocardiography (FE), as depicted in Figure 2a (further details in Section B.2). In parallel, time-varying flow rates were calculated based on blood flow velocity data acquired using spectral Doppler methods (Figure 2b; details in Section B.3). The anatomy and flow were thoroughly discussed and validated by the clinical contributors to this study (GC, LS, MS), with final verification by GS, before incorporating them into the novel iterative process shown in Figure 2c. This process involves finite element analysis (FEA) and an in-house Python code to estimate material properties and Windkessel (WK) parameters of the fetal aorta. This combined approach culminated in an FSI model that enabled personalized assessment of fetal aortic hemodynamics.

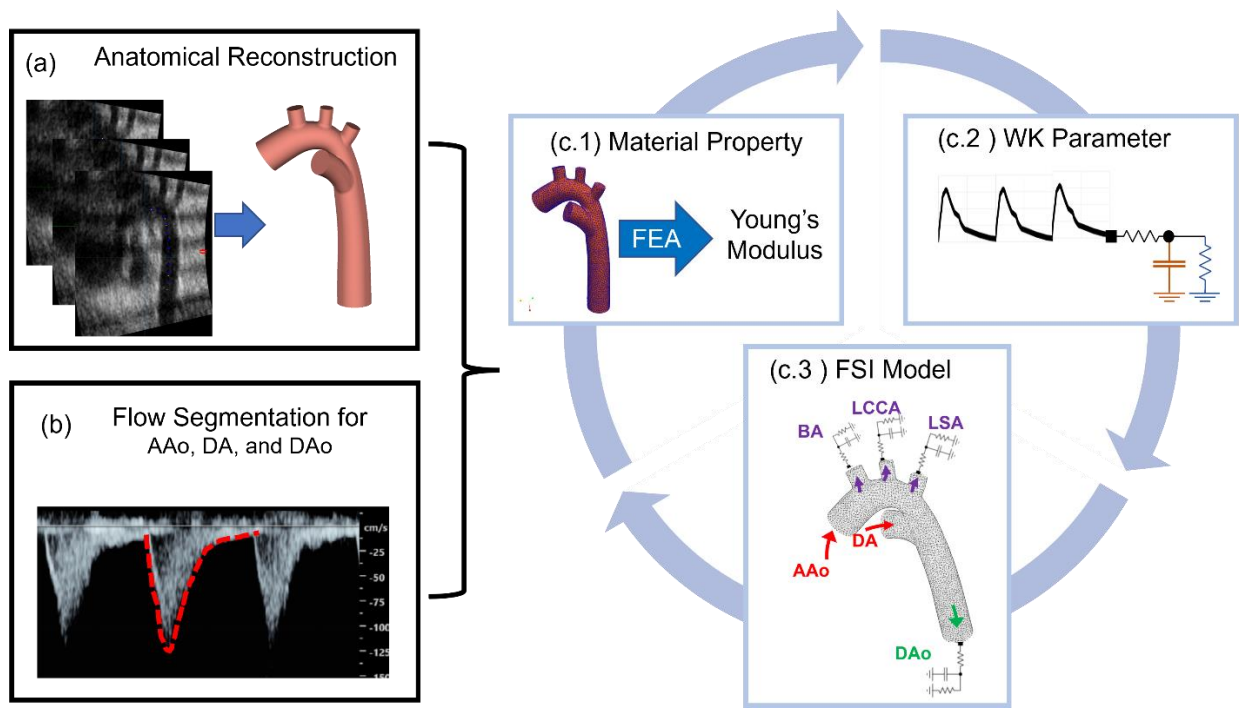

Figure 2. Proposed modeling pipeline for the fluid-structure interaction model. (a) The general steps used in this study start with anatomical reconstruction. Meanwhile, flow segmentation for the aortic arch, ductus arteriosus, and descending aorta is conducted, as shown in (b). (c) A sketch of the iterative approach proposed in this study that involves (c.1), a finite element analysis to estimate the bulk material property of the fetal aorta; (c.2) a Python code to tune automatically for parameters of the model, and (c.3), a fluid-structure interaction that reflects physiologically realistic pressure and flow inside the fetal aorta. AAo: ascending aorta; AoA: aortic arch; BA: brachiocephalic artery; DA: ductus arteriosus; DAo: descending aorta; FSI: fluid-structure interaction; LCCA: left common carotid artery; LSA: left subclavian artery; PA: pulmonary artery; WK: Windkessel.

### B.1 Study Subjects

The patients included in this study were retrospectively selected with the following criteria:

- Inclusion criteria: (i) Singleton pregnancy; (ii) gestational age (GA) > 20 weeks and < 40 weeks; (iii) no structural defect, heart failure, or arrhythmia in the fetus.

- Exclusion criteria: (i) Cardiac condition [e.g., fetal cardiac dysfunction and/or fetal hydrops, fetal arrhythmia, abnormal baseline heart rate (<110 bpm or > 160 bpm)]; (ii ) multiple pregnancies, and (iii) fetal growth restriction (<3rd centile).

Standard ultrasound systems (GE Voluson E-10) were used to acquire two-dimensional (2D), 3D, and Doppler FE. The long-axis acquisition was used to capture the center plane of the aortic arch with the ascending aorta (AAo), aortic arch, the descending aorta (DAo), the brachiocephalic artery (BA), the left common carotid artery (LCCA), and the left subclavian artery (LSA). The short-axis acquisition should exhibit the center plane of the 3-vessel view, including the aortic and ductal arch structures. The 2D FE images were also acquired. Doppler velocity tracing was recorded for the AAo, DA, aortic isthmus (AoI), and DAo. There was no clinical measurement in the LCCA, the LSA, and the BA, primarily because of small vessel sizes and challenges in designating an appropriate field of view.

**B.2  Anatomy Reconstruction and Flow Segmentation**

Three-dimensional FE acquisition generally includes approximately 30 slices spatially (with 0.5 mm thickness) and 7–13 time frames. The frame of the smallest volume of the fetal aorta (during end-diastole) was chosen for anatomical reconstruction using SimVascular [30]. First, the center plane of the fetal aorta was identified, highlighting the center plane of the aortic arch and its bifurcations to the LCCA, LSA, and BA. The next step was determining the bifurcation between the aortic arch and the DA after the centerlines of the aortic arch; the LCCA, LSA, BA, and DA were identified and then lofted, assuming their structures to be circular cylinders. The diameter of the lofted vessels was compared with the clinical measurements done in 2D FE, which is the current gold standard. All 2D FE measurements were verified by SG (the last author), who has over 20 years of clinical experience in acquiring and analyzing FE images and data.

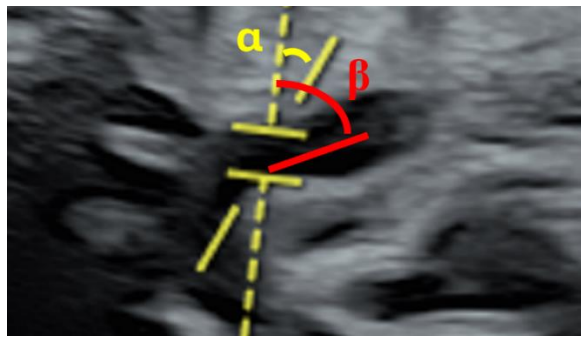

Figure 3 Angle correction of the velocity tracking

Doppler data were digitized and denoted as $V_{segmented}$. The ultrasound machine applied an initial angle correction (α) to the velocity tracking, as shown in Figure 3. However, this angle may differ from the actual angle (β) between the ultrasound beam and the true centerline of the vessel. This discrepancy primarily arises because of limitations in anatomical resolution when imaging fetal circulation [31, 32]. If the angle discrepancy (β-α) exceeded 20 degrees (introducing about a 10% error [33]), a secondary angle correction was applied to the velocity waveforms using the formula, $V_{corrected} = V_{segmented} \times (\cos\alpha/\cos\beta)$, consistent with the initial angle correction implemented by the ultrasound machine [31, 33]. The vessel flow rate (Q) was then calculated using the corrected velocity and vessel diameter: $Q = \pi r^2 V_{corrected}$, where $r$ is the radius of the vessel.

**B.3 Estimation of Material Property**

The bulk material property of a fetal aorta was estimated using a similar inverse method with an FEA [34, 35]. The process started with an initial guess of Young's modulus. A pulse pressure (ΔP) was applied inside the reconstructed fetal aorta (at its smallest volume). The vessel inflation function of SimVascular was used to deform the fetal aorta based on the current value of Young's modulus [36]. The vessel diameters of AoI in the deformed state were compared against clinical measurements. If the discrepancies were found to have exceeded 5%, Young's modulus was adjusted following the Newton-Raphson method, and another FEA based on the updated modulus was conducted. We adopted three assumptions commonly accepted by previous studies: (i) a homogeneous, linear elastic material model [19, 37]; (ii) the vessel wall thickness as 15% of the average radius of all vessel ends associated with the fetal aorta [25, 38]; and (iii) the Poisson ratio as 0.499 [37-39]. The challenge of deriving physiologically realistic material properties lies in selecting the appropriate pulse pressure to construct the traction vector, as detailed in Section B.6, which describes the iterative process.

**B.4 Fluid-Structure Interaction Modeling**

SimVascular has been used to generate the computational mesh, and the coupled momentum method has been used to conduct the FSI simulations [40]. In this method, a linearized kinematics Eulerian approach is adopted for monolithically coupling the fluid and solid domains. The fluid-solid meshes are kept fixed; therefore, the degrees of freedom of the vessel wall and the fluid boundary are identical, thus naturally satisfying the no-slip condition.

The incompressible Navier-Stokes equations were solved in the fluid domain:

$$\nabla \cdot v = 0, \quad \rho v_{,t} + \rho v \cdot \nabla v = -\nabla p + \nabla \cdot \tau \tag{1}$$

where $v$ represents the blood velocity vector, $\rho$ is the density of blood, $p$ stands for pressure, and $\tau$ is the viscous stress tensor defined as $\tau = \mu(\nabla v + (\nabla v)^T)$.

The vessel wall mechanics are approximated using a thin-walled structure assumption. Therefore, the solid domain is topologically defined by the same surface as the lateral boundary of the fluid domain. The elastodynamic equations, including boundary and initial conditions, for the vessel wall can be expressed as follows.

$$\rho^s u_{,tt} = \nabla \cdot \sigma^s + b^s \tag{2}$$

where $b^s$ is a body force per unit volume, and $\sigma^s(u)$ is the vessel wall Cauchy stress tensor.

In FSI simulations for cardiovascular modeling, Neumann boundary conditions (typically applied at outlets) may experience backflow because of either complete flow reversal (resulting from conservation of mass) or partial reversal caused by vortical structures near the boundary. This issue is particularly common when using the Windkessel model at the boundary [41-43]. Backflow can contribute to the numerical divergence of the simulation; therefore, backflow stabilization techniques are often implemented to mitigate this problem. In this study, we adopted the default backflow stabilization method provided by SimVascular, which introduces an additional convective term to the Neumann boundary condition [43]. Previous studies have shown that this stabilization technique offers a more robust and effective solution for handling flow reversals at Neumann boundaries compared to alternative methods, such as using Lagrange multipliers or constraining the velocity to be normal to the outlet [30, 44].

Inspired by Womersley's deformable wall theory, the coupled momentum method adopts a condition between the fluid and the solid domains: surface traction equality. The surface traction $t^f$ acting on the fluid lateral boundary because interaction with the solid is equal and opposed to the surface traction $t^s$ acting on the vessel wall due to the fluid: $t^f = -t^s$. Using a thin-wall approximation, the surface traction $t^s$ can be used to define a fictitious body force $b^s$ acting on the solid domain. Thus, on $\Gamma_s$ we have: $b^s = -t^f/h$, where h is the thickness of the vessel wall.

The linear solver employed was the generalized minimal residual method (GMRES), with a modified Newton-Raphson method applied for linearization. The convergence threshold for fluid-solid interaction was set at $10^{-6}$. The time-step size was determined as $T/800$, where $T$ represents the duration of the cardiac cycle. No turbulence modeling was involved because the Reynolds numbers based on the DAo for the subjects were 548±200. Blood was treated as a single-phase, Newtonian fluid using the previously published relation for fetal aortas [25]:

$$\mu = (1.15 + 0.075 \times GA)/1000 \ (kg/m\text{-}s) \quad (3)$$

A periodically stable solution is often achieved after the simulation completes two cardiac cycles. However, all cases were run for five cycles, and the data from the ending cycle were post-processed for reporting.

### B.5 Boundary Conditions

Womersley velocity profiles were applied to the AAo and DA along with imposing 3-element WK models on BA, LCCA, and LSA, and DAo, as shown in Figure 4.

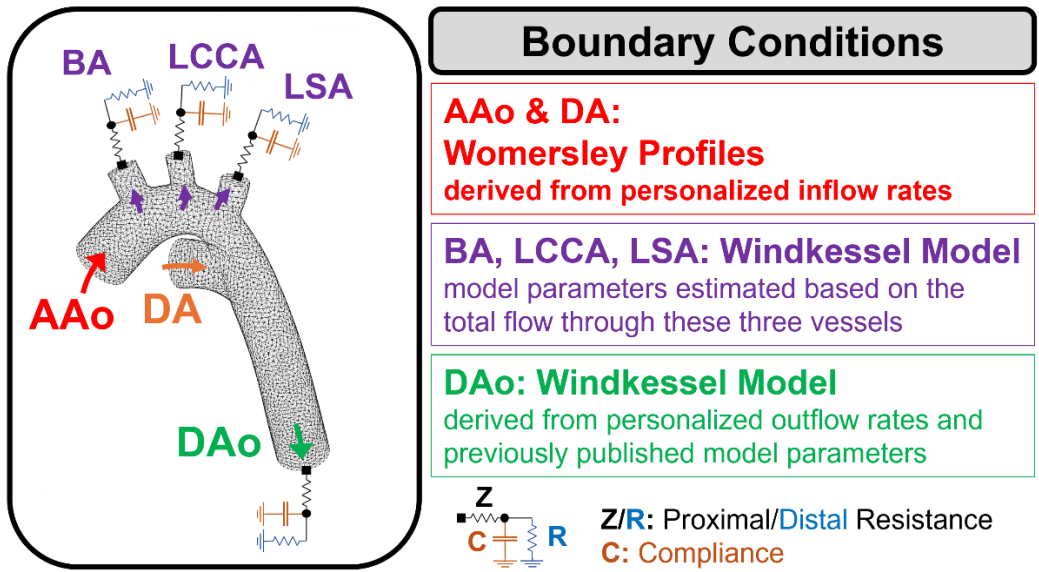

Figure 4. The mesh and boundary conditions of the final fluid-structure interaction model. AAo: ascending aorta; BA: brachiocephalic artery; DA: ductus arteriosus; DAo: descending aorta; LCCA: left common carotid artery; LSA: left subclavian artery.

An in-house Python code, PyAuto3WK, was developed to automate the estimation of parameters for the WK model. This code takes the *in vivo* flow measurement as input and generates the proximal and distal resistance (Z and R, respectively) and compliance as outputs. Pattern search optimization is employed to automatically seek the values of Z, R, and compliance (C) that reconcile the output pressures with the target mean and pulse pressure while ensuring that proximal resistance remains below 10% of the distal resistance (Z < 0.1R). This constraint between Z and R reflects a physiological state validated by clinical measurements [45] and is widely accepted in numerical studies [36, 46-49].

### B.5.1 Windkessel Parameters for the Descending Aorta

Obtaining the WK parameters for the DAo is relatively simple. The input flow is acquired through in vivo Doppler measurements. Because invasive pressure measurements are unavailable, we relied on previously documented mean and pulse pressure values for the DAo in fetuses, which depend on GA [25, 39]. These values are denoted as $\overline{P_{DAo\_GA}}$ and $\Delta P_{DAo\_GA}$, respectively.

$$\overline{P_{DAo\_GA}}(mmHg) = 0.87 \times GA + 10.33 \tag{4}$$

$$\Delta P_{DAo\_GA}(mmHg) = 0.39 \times GA + 13.44 \tag{5}$$

### B.5.2 Windkessel Parameters for the Three Branches

Estimating the WK parameters for the BA, LCCA, and LSA (hereafter referred to as the "three branches") presents three main challenges:

1. *Lack of in vivo flow measurements for the BA, LCCA, and LSA.* To address this issue, we estimate their total flow rate using $Q_{3b} = Q_{AAo} + Q_{DA} - Q_{DAo}$.

2. *Absence of in vivo pressure measurements or reported literature values for the three branches.* Consequently, we initially used literature-reported pressures at the DAo as a starting point [equations (4) and (5)]. Subsequently, we refined the pressure estimation at the three branches through an iterative process proposed in this study (details provided in Section B.6).

3. *No validated method for "splitting" the WK parameters for the BA, LCCA, and LSA in the fetal aorta.* Upon calculating the total WK parameters for the three branches, we need to "split" them accordingly to the BA, LCCA, and LSA.

    a. The "splitting" algorithm for the total resistance ($R_{total}$) was usually based on the radius of the branches:

    $$R_i = R_{total} \frac{\sum r_i^m}{r_i^m} \tag{6}$$

    where $R_i$ and $r_i$ (i = 1, 2, 3) are the resistance and radius of the three branches, respectively. The coefficient *m* in the power law varies based on different theories. The "square law" (m = 2) is derived from curve-fitting clinical data in various scenarios, such as the internal carotid artery [50] and the major branches of the adult aorta [51]. The "cube law" (m = 3) adheres to Murray's law,

which dictates the physiological principle of minimum work [52, 53]. The "fourth-order law" follows the Hagen-Poiseuille law, where flow resistance is proportional to the square of the vessel area. This law has been applied in the 0D hemodynamic assessment of the fetal aorta [25].

  b. The total compliance ($C_{total}$) has a similar "splitting" algorithm:

$$C_i = C_{total} \frac{r_i^m}{\sum r_i^m} \tag{7}$$

where $C_i$ (i = 1, 2, 3) is the compliance of the three branches. The "square law" (m = 2) has been used for the adult aorta [33], and the "cube law" (m = 3) has been used in 0D flow modeling of the fetal aorta [24].

No prior studies have addressed the optimal power law for the 3D hemodynamic assessment of the three branches of the fetal aorta. Consequently, this study investigates and compares the effectiveness of these laws for this purpose.

### B.6 Iterative Process

The goal of this iterative procedure is to achieve physiologically reasonable pressure within the fetal aorta. Once again, the pressure drop attributed to the inclusion of the DA flow (as explained in the Introduction) renders it impractical to rely on $\overline{P_{DAo\_GA}}$ and $\Delta P_{DAo\_GA}$ for obtaining physiologically reasonable material properties (described in Section B.2) and WK parameters for the three branches (described in Section B.5).

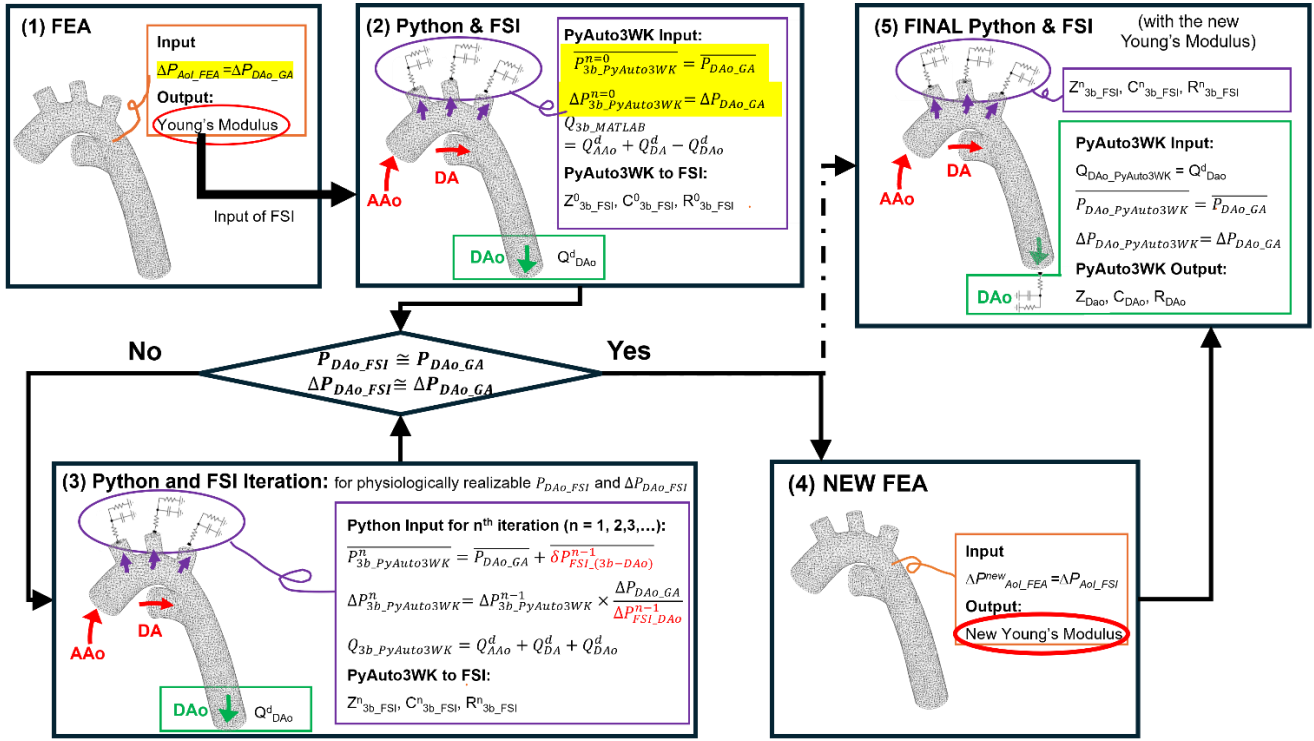

Figure 5. Workflow of the iterative process proposed in this study. In steps 1 and 2, the highlighted equations utilize pressure data from the literature for the descending aorta as an initial approximation for the pressures at the aortic isthmus and the three branches. Although this initial approximation is not physiologically accurate, it serves as a starting point for the subsequent steps. AAo: ascending aorta; AoI: aortic isthmus; C: compliance; DA: ductus arteriosus; DAo: descending aorta; FEA: finite element analysis; FSI: fluid-structure interaction; GA: gestational age; PyAuto3WK: in-house Python code; R: distal resistance; Z: proximal resistance.

Therefore, we devised an iterative process comprising five steps, delineated in Figure 5. These steps can be divided into three stages:

- *The initial stage* encompasses steps 1 and 2 depicted in Figure 5. During this step, we initially employed $\Delta P_{DAo\_GA}$ as the first guess of pulse pressure at the AoI for the FEA ($\Delta P_{AoI\_FEA}$ in Figure 5) to approximate a bulk material property of the fetal aorta. In step 2, we utilized the pressure at DAo from the literature [$\Delta P_{DAo\_GA}$ and $\overline{P_{DAo\_GA}}$, equation (4) and (5)] in the PyAuto3WK code to estimate the WK parameter for the three branches. It is noteworthy that, in this step, the FSI setup differs slightly from the one illustrated in Figure 4. The only difference lies in the boundary condition at the DAo, which is an outflow velocity condition based on the *in vivo* measurement rather than on a WK model.

- *The iterative stage* comprises step 3 in Figure 5, which involves multiple iterations between the FSI model and the PyAuto3WK code. For each iteration *n*:

- The simulated pressures at DAo ($\overline{P_{DAo\_FSI}}$ and $\mathit{\Delta}P_{DAo\_FSI}$) and at the three branches $\overline{P_{3b\_FSI}}$ were first obtained from the FSI model of iteration *n-1*.
  - $\overline{P_{DAo\_FSI}}$ was used to calculate the pressure drop from the three branches to DAo: $\overline{\delta P_{FSI\_(3b-DAo)}} = \overline{P_{FSI\_3b}} - \overline{P_{FSI\_DAo}}$, which was then used to update the mean pressure of the three branches in our PyAuto3WK code.
  - $\mathit{\Delta}P_{DAo\_FSI}$ corrected the pulse pressure of the three branches in the PyAuto3WK code.
- The iteration ceased when the stimulated mean and pulse pressure at DAo agreed well with the literature data (within a 5% of error margin).

- *The final correction stage* includes steps 4 and 5 in Figure 5. By the conclusion of step 3, the pressures at the three branches and AoI could be remarkably different from their initial value used in *the initial stage* (steps 1 and 2). Consequently, it is appropriate to use the updated $\mathit{\Delta}P_{AoI\_FSI}$, which is more physiologically acceptable than $\mathit{\Delta}P_{DAo\_GA}$, to re-estimate the bulk material property of the fetal aorta. Employing this updated material property and the last set of WK parameters for the three branches from step 3, another FSI model is constructed in step 5. This FSI mirrors the one illustrated in Figure 4, where the DAo also uses a WK model, whose parameters were directly calculated from the PyAuto3WK code based on in vivo measured DAo flow and pressures at DAo from the literature [equations (4) and (5)].

### B.7 Statistical Analysis

The primary outputs of the FSI model include velocity, pressure, and wall shear stress (WSS). Statistical analyzes were performed using IBM SPSS Statistics (IBM, Inc., Aramark, NY, USA). The metrics were first analyzed for normality using the Shapiro-Wilk test. Depending on normality results, either a Pearson or a Spearman correlation test was employed to ascertain the correlation between the FSI-derived metrics and the *in vivo* measurements.

## C. Results

### C.1 Patient Selection

The personalized anatomies for all subjects have been successfully reconstructed, as shown in Figure 6. All anatomies were plotted on the same dimensional scale to facilitate the visualization of their size differences.

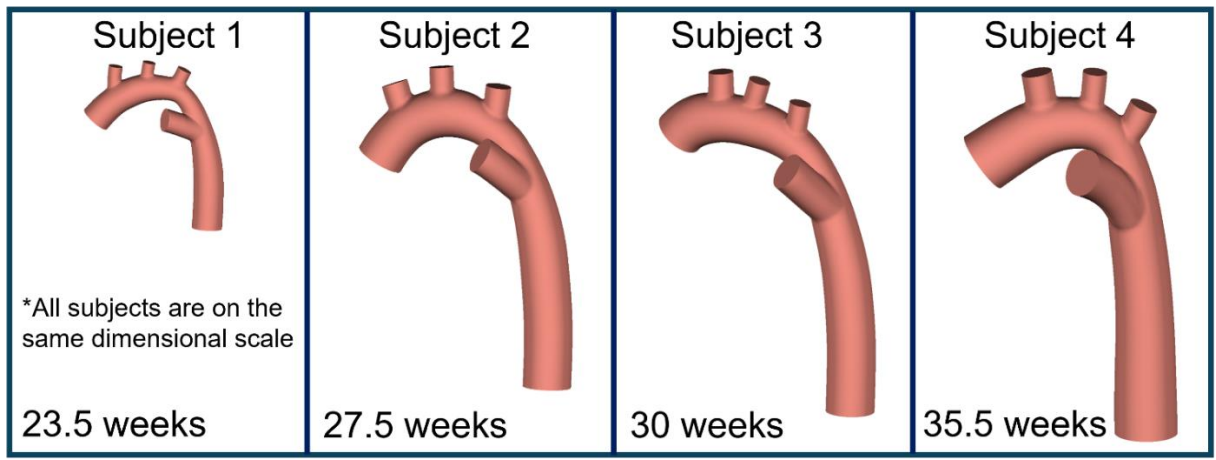

Figure 6. Personalized anatomies for all subjects, with the corresponding gestational age.

These individuals cover the typical GA range that was feasible for FE acquisition. In Figure 7, the vessel size determined by 2D FE and FSI was also compared, and good agreements were obtained. Table 1 shows the flow information of AAo, DA, and DAo. All vessels presented an increasing flow rate with GA. The highest Reynolds numbers in each subject occur at DAo; none exceeds the critical Reynolds number for a circular tube, i.e., 2000.

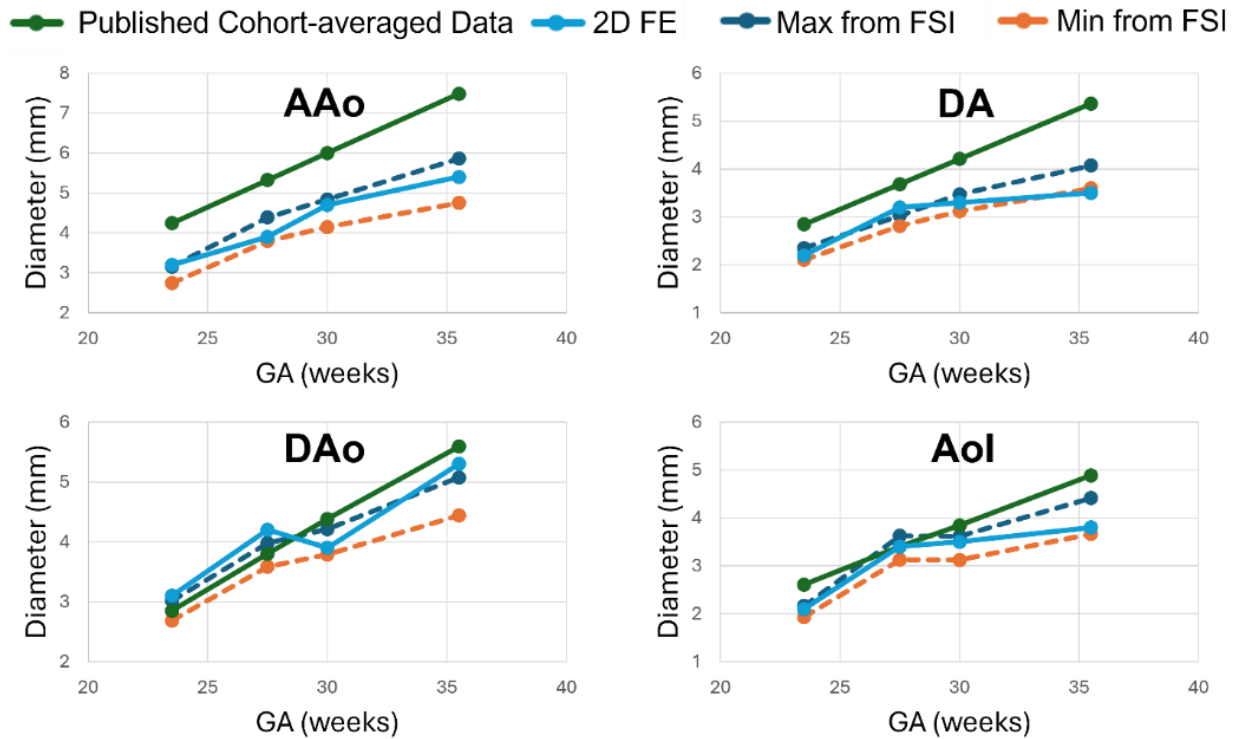

Figure 7. Comparisons of vessel sizes between two-dimensional fetal echocardiography, fluid-structure interaction, and published cohort-averaged data [25]. 2D: two-dimensional; AAo: ascending aorta; AoI: aortic isthmus; DA: ductus arteriosus; DAo: descending aorta; FE: fetal echocardiography; FSI: fluid-structure interaction; GA: gestational age.

Table 1. Summary of flow rate and Reynolds number for ascending aorta, ductus arteriosus, and descending aorta.

| | AAo | | DA | | DAo | |
|---|---|---|---|---|---|---|
| | Q (mL/min) | Re | Q (mL/min) | Re | Q (mL/min) | Re |
| Subject 1 (23.5 weeks) | 134 | 310 | 57.4 | 185 | 112 | 268 |
| Subject 2 (27.5 weeks) | 257 | 417 | 196 | 376 | 342 | 551 |
| Subject 3 (30 weeks) | 330 | 458 | 208 | 419 | 413 | 730 |
| Subject 4 (35.5 weeks) | 589 | 627 | 196 | 321 | 594 | 642 |
| AAo: ascending aorta; DA: ductus arteriosus; DAo: descending aorta; Re: Reynolds number | | | | | | |

## C.2 Mesh-independence Study

The mesh-independence study was conducted based on subject 4. The study started with a $D_{avg}/5$ edge size, where $D_{avg}$ is the average diameter of inflow boundaries, i.e., DAo and DA. Every refinement reduced the edge size by 20%, resulting in a ~50% reduction in mesh cell volume. Figure 8 shows that the $D_{avg}/7.94$ produced mesh-independent results in velocity (less than 3% changes) [54, 55], the metric of interest for this study. Therefore, this mesh was chosen for all simulations, resulting in around 0.25 million tetrahedral cells per mesh.

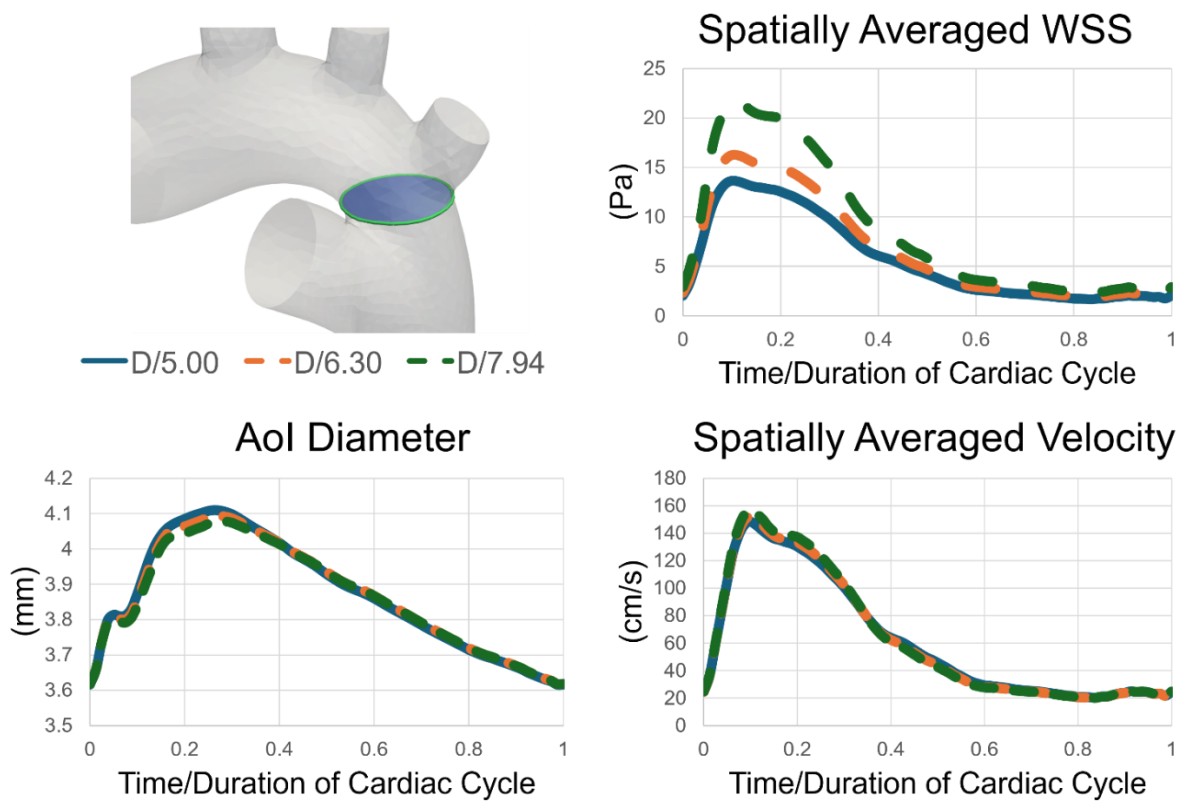

Figure 8. Results of mesh independence study for a representative subject. AoI: aortic isthmus; D: diameter; WSS: wall shear stress.

## C.3 Windkessel Parameters for Three Branches

Section B.5 introduces six parameter-splitting algorithms (m = 2,3,4 for resistance in equation 6 and m = 2,3 for compliance in equation 7). The only approach employed in fetal aortic simulations was the one characterized by $R_i \propto r_i^4$, $C_i \propto r_i^3$ (m = 4 for equation 6 and m = 3 for equation 7) [24] . However, this study involved a simplified computational model instead of a high-fidelity 3D simulation. Other splitting algorithms have also been widely used in various other studies. Therefore, we explored the differences between the methods characterized by $R_i \propto r_i^4$, $C_i \propto r_i^3$ and by $R_i \propto r_i^2$, $C_i \propto r_i^2$, given that the coefficients of other algorithms lie between these two extremes. Figure 9 illustrates the results, showing that the majority of the variances in pressure and WSS are concentrated on the three branches. The impact of varying the splitting algorithm on the AoI is negligible (<3%). Therefore, we used the $R_i \propto r_i^4$, $C_i \propto r_i^3$ for all simulations in this study.

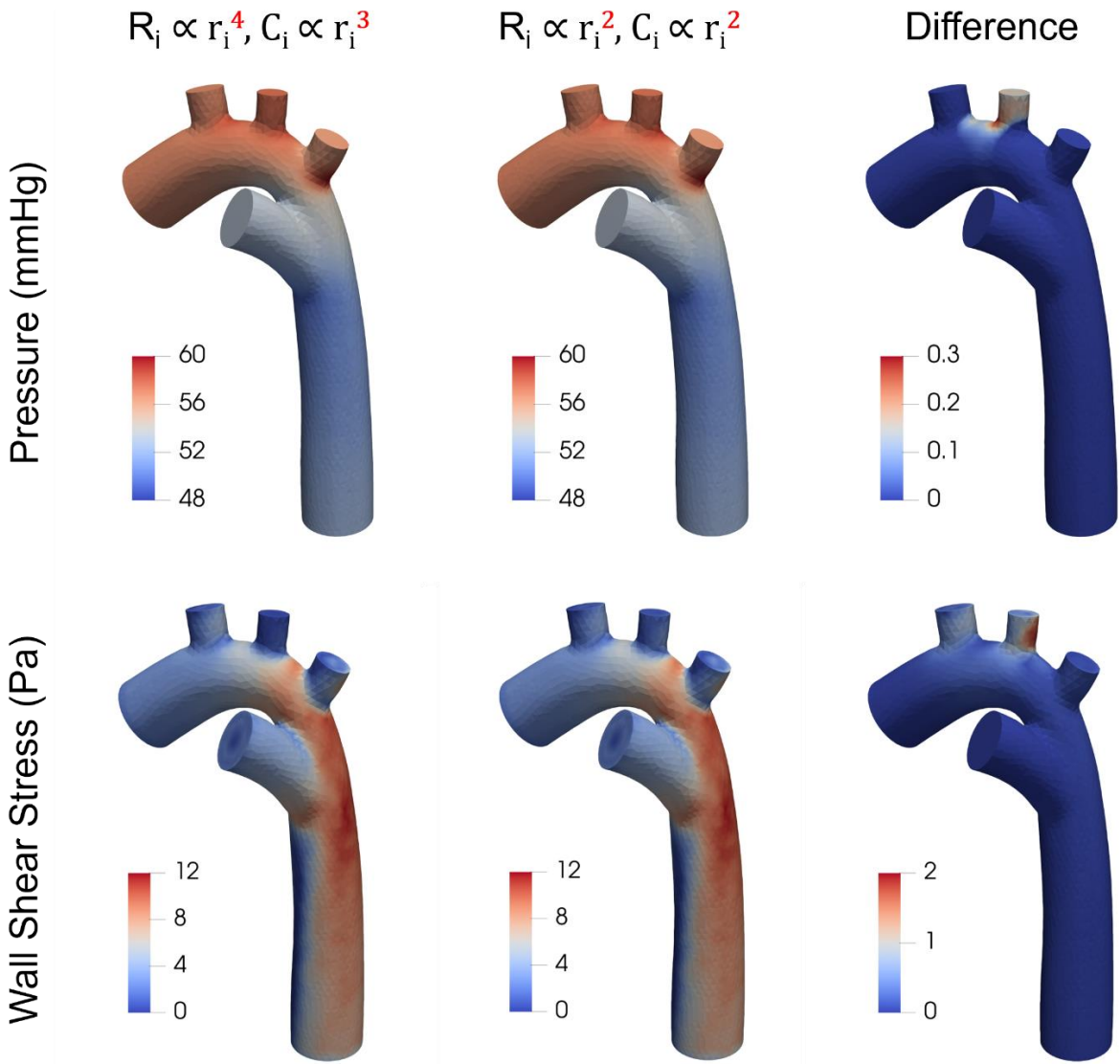

Figure 9. A comparison of parameter-splitting algorithms applied to the three branches for a representative subject. The left column pertains to the method with m = 4 for equation 6 and m = 3 for equation 7, and the middle column corresponds to the one with m = 2 for equation 6 and m = 2 for equation 7.

### C.4 Convergence of the Iterative Process

Figure 10 illustrates the criteria for stopping the iterative process: (i) the $Q_{DAo}$ from FSI matches the in vivo measured value (R ≥ 0.95) and (ii) the pulse pressure of DAo ($\Delta P_{DAo}$) from FSI matches the $\Delta P_{GA}$ (error < 5%).

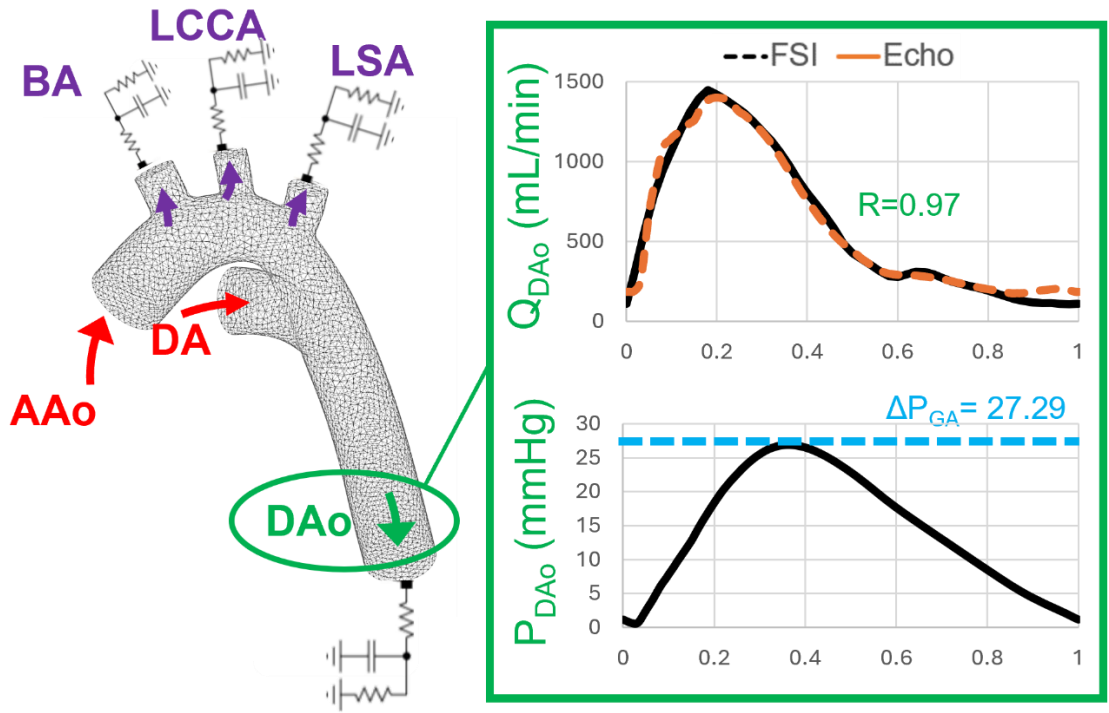

Figure 10. Model verification for a representative subject. AAo: ascending aorta; BA: brachiocephalic artery; descending aorta; DA: ductus arteriosus; DAo: descending aorta; Echo: echocardiography; FSI: fluid-structure interaction; LCCA: left common carotid artery; LSA: left subclavian artery.

It is worth noting that the iterative stage does NOT involve any trial-and-error steps. Therefore, it usually takes three iterations before convergence; an example is shown in Figure 11. The value of $\Delta P_{DAo\_FSI}$ is usually lower than the anticipated value before the convergence, and $\overline{P_{FSI}}$ could fluctuate between the target values.

Figure 10 also shows that using the pressures from literature, instead of an initial guess of zero, facilitates the convergence of the iterative process. Also, the difference of $\overline{P_{FSI}}$ and $\Delta P_{DAo\_FSI}$ from the three branches to DAo are 3.2–14.7% and 3.6–41.1% based on the value at the Dao, respectively.

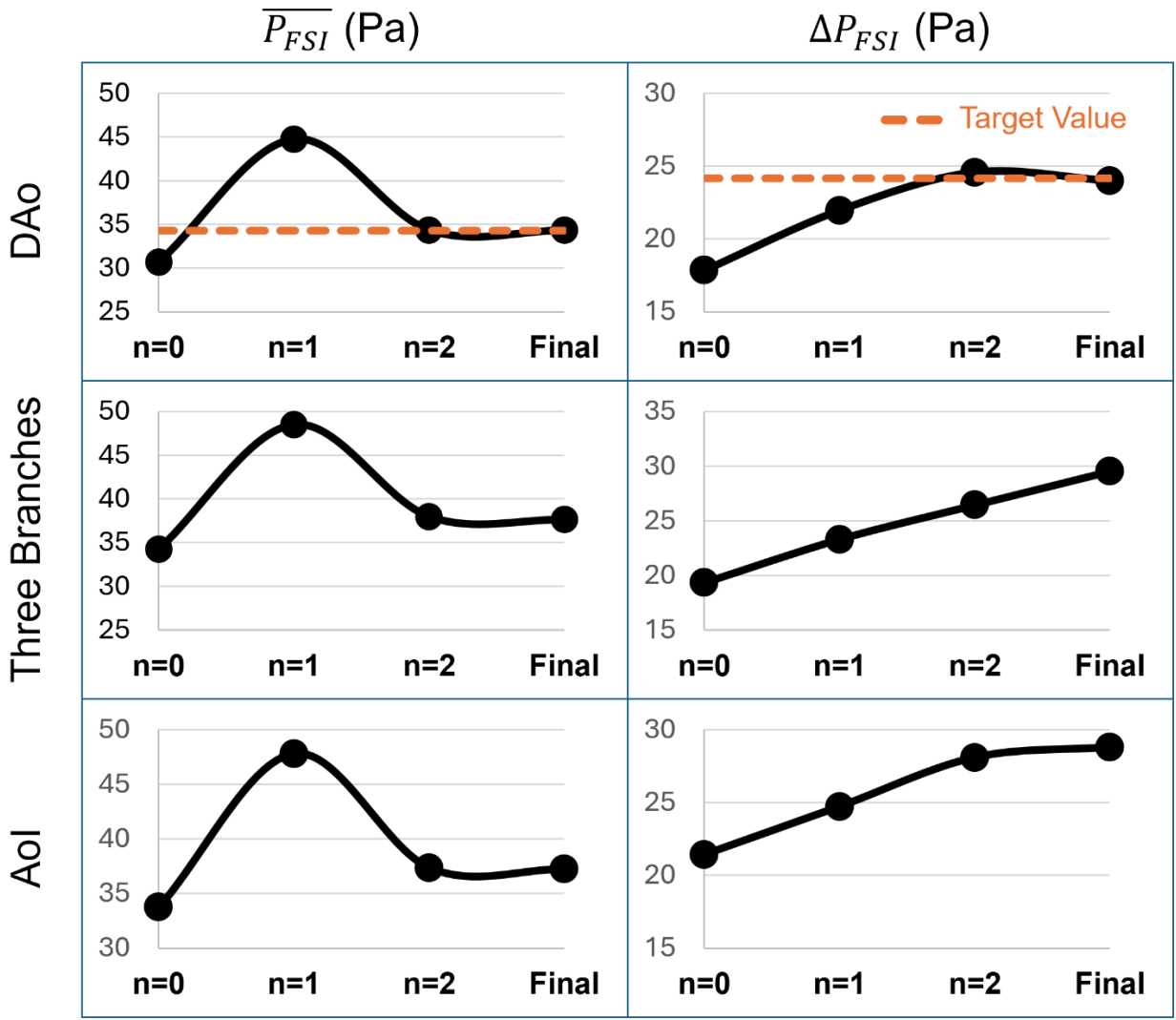

Figure 11. Pressure histories of a representative subject during the iterative stage. Here, n = 0 corresponds to step (2) in Figure 5; n ≥ 1 is linked to step (3) in Figure 5; and "Final" denotes step (5) in Figure 5. AoI: aortic isthmus; DAo: descending aorta; FSI: fluid-structure interaction.

### C.5 Validation

The criteria of a successful validation were that (i) the diameter of the AoI matches the in vivo measured value (error < 5%) and (ii) the velocity at AoI reasonably agrees with the in vivo measurement (R > 0.95). Figure 12 demonstrates the validation results of the proposed FSI model in four subjects. They all successfully met the two criteria.

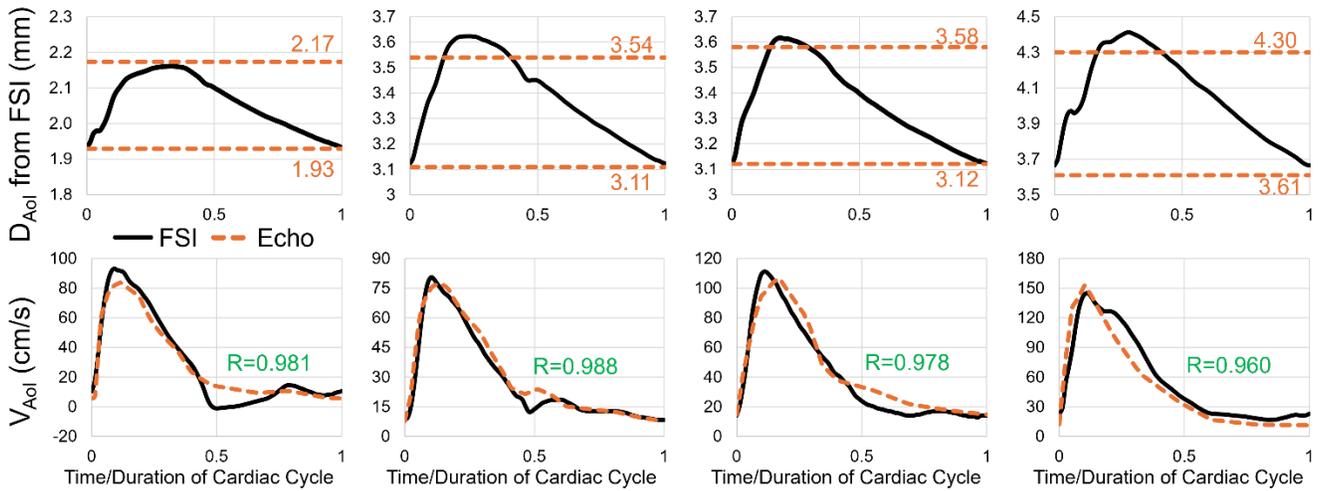

Figure 12. Comparisons for the velocity and diameter at the aortic isthmus for all subjects. $D_{AoI}$: diameter of the aortic isthmus; Echo: echocardiography; FSI: fluid-structure interaction; $V_{AoI}$: velocity at the aortic isthmus.

## C.6 Biomechanics of Fetal Aorta

Figure 13 illustrates the pressure and particle tracking results for all subjects. The pressure contours explicitly depict the remarkable pressure drop from the AAo to the DAo. The particle line indicates that AAo flow bifurcates into four streams: three of them flow to BA, LCCA, and LSA; the other one merges with the DA flow and exists through the DAo.

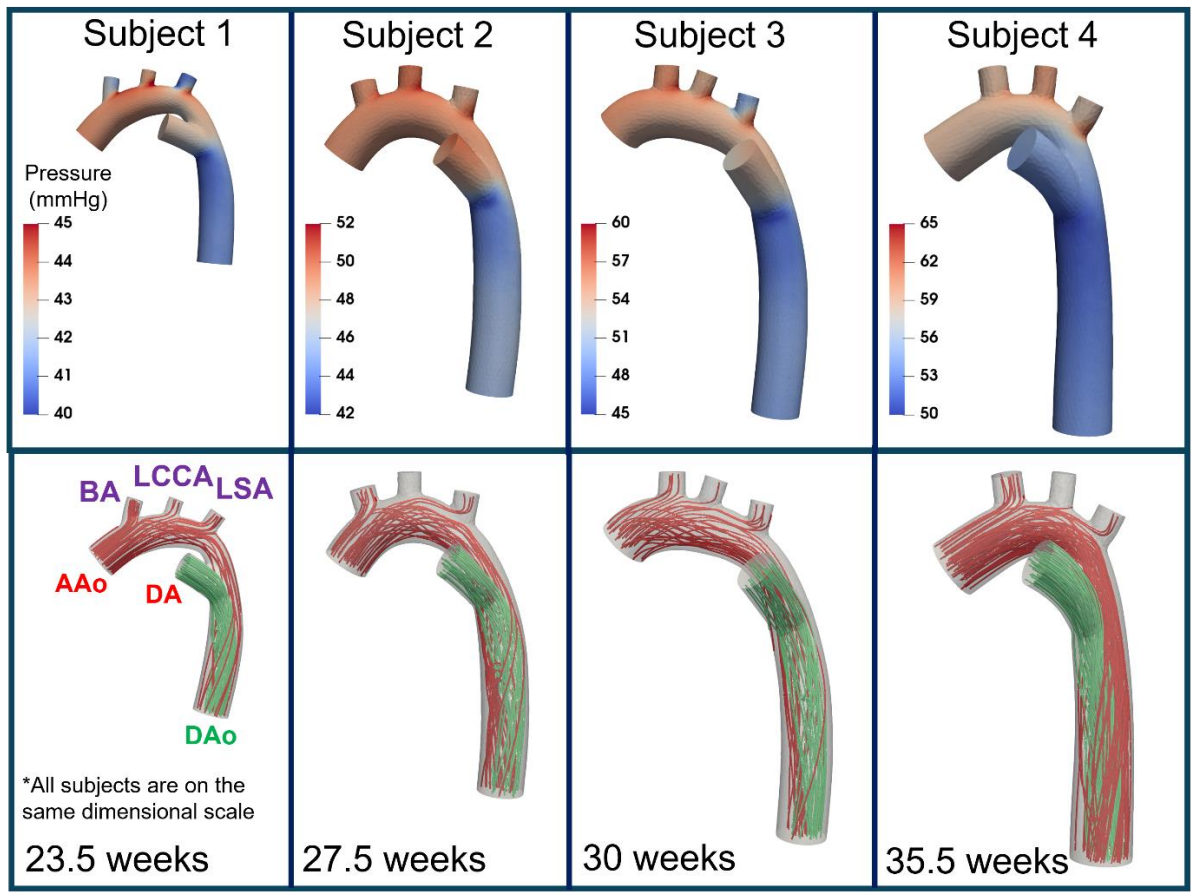

Figure 13. Pressure and particle path (colored by velocity) for all subjects. AAo: ascending aorta; BA: brachiocephalic artery; DA: ductus arteriosus; DAo: descending aorta; LCCA: left common carotid artery; LSA: left subclavian artery.

Table 2 summarizes the material property and diameter changes for all subjects, showing the AoI diameters obtained in good agreement between the FEA and the *in vivo* measurements.

Table 2. Material properties and vessel changes of all subjects.

|  | Young's Modulus (MPa) | $D_{AoI}$ in mm (min-max) | |
|---|---|---|---|
|  |  | Simulation | in vivo |
| Subject 1 | 0.165 | 1.93-2.16 | 1.93-2.17 |
| Subject 2 | 0.185 | 3.12-3.63 | 3.11-3.54 |
| Subject 3 | 0.2 | 3.12-3.62 | 3.12-3.58 |
| Subject 4 | 0.15 | 3.66-4.41 | 3.61-4.30 |
| $D_{AoI}$: diameter of the aortic isthmus; MPa: Megapascal; | | | |

## D. Discussion

### D.1 Patient Selection

This study meticulously chose four fetal aortas that span a broad spectrum of GAs, featuring remarkably diverse geometrical dimensions (illustrated in Figures 6 and 7) and flow conditions (demonstrated in Table 1). Both the geometry and flow increase with GA, with the rate of increase closely aligning with that reported in previous studies [19]. The highest Reynolds number is observed in DAo, as detailed in Table 1. However, no evidence of turbulence was observed because of the relatively low Reynolds numbers across all cases.

### D.2 Model Development, Verification, and Validation

The primary difference between the fetal aorta and the pediatric/adult aorta is the presence of the DA. This vessel complicates both the anatomy and FSI modeling of fetal aortas. In addition, the in vivo measurement is limited because the pressure is unavailable for all vessels. Previous articles in the literature only reported the mean and pulse pressure in the DAo. The remarkable pressure drop from the AAo to the DAo, highlighted in Figures 11 and 13, introduces two notable "discontinuities" in developing our workflow:

1. The determination of the fetal aorta's material properties necessitates knowing the pulse pressure at the AoI, which is considerably higher than that at the DAo by as much as 41.1%.
2. The calculation of the WK parameters for the three branches (BA, LCCA, and LSA) requires both the mean and pulse pressures at the AAo. Whereas the difference in mean pressure is minimal (around 3%) in some subjects, it can reach up to 14% in others.

All these "discontinuities" represent challenges for personalized FSI of the fetal aorta. However, our workflow takes less than three iterations to achieve all verification requirements. Moreover, we conducted the validation in a more rigorous way, because the correlation coefficients (in Figure 12) factor in all errors during a cardiac cycle. Previous studies validated their model concerning single velocity values, e.g., peak systole.

### D.3 Hemodynamics in the Fetal Aorta

In addition to the pressure drops across the AoI, the presence of the DA led to flow confluence, as shown in Figure 13. The remainder of the AAo flow through the AoI merges with the DA flow and generates a helical flow. The direction of flow swirling was clockwise downward to the DAo, similar to the flow within the DAo of children and adults [56, 57]. However, the helical flow observed in this study stems from flow confluence near the DA compared to the flow within children and adults, primarily driven by the twisting motion of heart contraction and aortic valve biomechanics. Nevertheless, the helical flow plays many positive physiological roles, including facilitating blood flow transport, enhancing nutrient transport from the blood to the arterial wall, and reducing the accumulation of atherogenic low-density lipoproteins on the endothelial surface of arteries [57].

Due to the absence of precise flow measurements for the three branches (BA, LSSA, and LCA), we assessed the robustness of applying fluid dynamics theories and clinical insights to flow splitting, as discussed in Section B.5. A specific method ($R_i \propto r_i^4$, $C_i \propto r_i^3$, depicted in Figure 9) has previously been employed in the 3D flow modeling of the fetal aorta. Yet, the influence of various flow-splitting algorithms on the 3D hemodynamic evaluation of the fetal aorta has not been scrutinized. Our findings indicate that the primary effect of these algorithms is observed in the three branches, the impact of which appears to be minimal on their downstream areas, such as at the AoI. However, our investigation was limited to a single fetus. A study involving a larger group of subjects is warranted for a comprehensive understanding, especially if the measurements in the three branches are significant.

### D.4 Aortic Stiffness

Ventricular-arterial coupling, a clinically important determinant of cardiovascular performance, measures interactions between the heart and the arterial system [58-61]. The proposed flow model produced great agreements regarding vessel deformation when we compared the AoI diameter obtained by FSI simulation and that obtained by clinical measurement. Our method represents the *first-of-its-kind* approach to estimating the material properties of the fetal aorta, gaining acceptability from prior simulation studies on fetal circulation [12, 13, 62]. Aortic stiffness is a crucial metric in the calculation of ventricular-arterial coupling. However, measuring in vivo aortic stiffness in fetal aortas is challenging, and previous studies adopted simplified estimations. The uncertainty of the stiffness measurements resulted in conflicting trends between the aortic stiffness in fetuses and GA. Some studies demonstrated a decrease in aortic stiffness with GA [23, 39], and others reported an increasing trend [59, 60]. A few recent studies support the decreasing trend using indirect evidence. Miyashita *et al.* revealed a negative relationship between pulse-wave velocity and GA and suggested that pulse-wave velocity was positively correlated with vessel stiffness [61]. Zhong *et al.* reported a decrease in the aortic strain as GA increases, alluding to the fact that the strain was significantly negatively associated with stiffness [58]. Our inversed method was much more high-fidelity than the simplified theoretical models used in previous studies. Interestingly, the information presented in Table 2 indicates a mixed pattern in aortic stiffness relative to GA: It shows an increase prior to 30 weeks and a decrease thereafter. However, the limited size of the sample makes it necessary to approach the generalization of these findings with caution, suggesting the need for further research.

**D.5 Clinical Implications**

Computational flow models offer a cost-effective means to analyze blood flow and hemodynamics, providing critical insights into the underlying mechanisms of cardiovascular and cerebrovascular diseases [63-65]. The knowledge and tools developed through these models can be applied to improve both diagnosis and prognosis [66, 67]. Additionally, advancements in model-based diagnostics have opened avenues for commercialization, with several start-up companies receiving FDA approval for clinical applications. The combined market value of these companies is estimated at $2.4 billion [68]. Moreover, the predictive capabilities of computational models have been utilized to design, test, and optimize medical devices and therapies for cardiovascular and cerebrovascular diseases[69-74], making them valuable tools for preclinical testing that can subsequently guide the design of clinical trials [75-78]. Finally, computational flow models are powerful tools for predicting the efficacy of proposed surgical interventions or medical devices based on pre-operative anatomical models (also known as surgical or treatment planning) [79-83]. This application of advanced CFD analyzes increases confidence in achieving desired surgical outcomes, prevents potential complications, shortens operative times, and minimizes the need for additional procedures [66, 81, 84-86].

In the context of fetal aortic conditions, the FSI model offers a cost-effective solution for obtaining high-fidelity measurements of WSS in fetal circulation - a crucial flow parameter that is challenging to capture using current imaging techniques. WSS plays a pivotal role in vessel dilation, remodeling, and recurrent coarctation in patients with repaired CoA [87-94]. As an innovative flow metric, WSS might distinguish between healthy and coarcted aortas, leading to advancements in the prenatal diagnosis of CoA, which remains one of the most frequently overlooked CHDs [6, 7]. Additionally, the FSI model could be used to assess the feasibility and safety of non-invasive therapeutic strategies for fetal CoA, such as maternal hyperoxygenation [95-97].

## E. Limitations

The present study has several limitations. The primary limitation is related to the material model used. The FSI model employed a homogeneous, linear elastic material model with a constant wall thickness. While these simplifications have been commonly used in previous studies of both fetal and adult aortas [19, 25, 37, 38], they do not fully capture the complex mechanical behavior of arterial tissue. Additionally, there is no direct validation of the assumed material properties against *ex vivo* measurements of fetal aortas. Although obtaining such measurements from healthy fetuses presents significant technical and ethical challenges, the good agreement between the simulated FSI results and clinically measured vessel dimensions and flow velocities provides support for the rationale and accuracy of this approach in estimating fetal aortic material properties. Nonetheless, future research should focus on developing more sophisticated material model [98, 99] and validating these models through ex vivo measurements where feasible.

The second limitation involves the boundary conditions, which were simplified using Womersley profiles at the inlets and a Windkessel model at the outlets. While this configuration is widely accepted in cardiovascular simulations [100-103], recent studies increasingly recommend using patient-specific velocity profiles that can only be accurately captured using phase-contrast magnetic resonance imaging [54, 104-106]. Despite the difficulties in obtaining such data for fetal circulation, research has shown that patient-specific velocity profiles are essential for accurately assessing localized WSS when compared to idealized profiles [54, 104, 105, 107, 108]. However, since this study does not focus on localized WSS, using idealized profiles could be acceptable. Similarly, due to the difficulties in acquiring patient-specific data for LCCA, LSA, and BA, a Windkessel model was implemented at these boundaries, and the proposed iterative process was used to achieve physiologically realistic flow distributions. As imaging techniques for fetal circulation advance, future studies should explore the effects of different boundary conditions to improve the precision of hemodynamic simulations.

Third, we calculated the flow rate for the three branches using the formula $Q_{3b}$ = AAo + DA – Dao. Although this approach is appropriate for computing average flow rates, it may introduce mass imbalances in instantaneous flow measurements due to residual volumes caused by vessel deformation. However, since instantaneous flow was used only to estimate the RCR values for the branches, and not directly applied to the FSI model, the mass imbalance has minimal impact. Furthermore, the WK model maintains mass conservation when integrating the RCR values into the FSI simulation.

Fourth, the computational mesh used in this study was designed to ensure mesh-independent results for velocity but not for WSS. Given that velocity was the primary metric of interest, this mesh configuration was deemed sufficient. Nevertheless, future studies focusing on WSS will require a more refined mesh to ensure an accurate assessment of shear stress.

Lastly, this study has a limited sample size. Although this is a constraint, it is acceptable in the context of previous validations, most of which involved even smaller cohorts [37, 109-112]. Future studies should include larger sample sizes to validate the generalizability of the findings.

## F. Conclusions

We proposed a novel personalized flow modeling paradigm utilizing FSI, which is time-efficient and capable of simulating high-fidelity hemodynamics of the fetal aorta. To date, this is *the first* paradigm that considers the personalized anatomy, flow, and material properties of the fetal aorta. It provides new evidence of the high-fidelity hemodynamics and biomechanics of the fetal aorta. More importantly, this paradigm was successfully implemented and rigorously validated against clinically measured dimensions and velocities over a cardiac cycle in four fetuses representing a range of GAs. Thus, this paradigm represents an important step toward developing a tool to understand the pathogenesis of CHDs related to the fetal aorta and to support clinical applications such as improving diagnosis and prognosis and exploring non-invasive therapeutic strategies for these CHDs.

## G. Acknowledgments

We thank the developers and contributors of the following open-source software used in this study, including SimVascular (https://simvascular.github.io/) for numerical simulation, RadiAnt DICOM Viewer (https://www.radiantviewer.com) for handling DICOM images, and 3D Slicer (https://www.slicer.org/) for refining reconstruction. We also extend our gratitude to the members of the Artificial Intelligence and Modeling Laboratory for Cardiovascular Diseases (AIMCardio Lab), including postdoctoral fellows, graduate students, undergraduates, and high school students, for their support in flow segmentation, anatomical reconstruction, method development, and document editing and refinement for this study.

## H. Disclosures

All authors declare that they have no conflicts of interest.